\documentclass{PoS}

\usepackage{amsmath}

\usepackage{txfonts}

\usepackage{float} 
\usepackage[caption=false]{subfig} 

\let\OLDthebibliography\thebibliography
\renewcommand\thebibliography[1]{
  \OLDthebibliography{#1}
  \setlength{\parskip}{0pt}
  \setlength{\itemsep}{-2pt}
}

\usepackage{lineno}

\newcommand{\ie}{{\it i.e.}}

\usepackage{xspace}
\usepackage{comment}

\newcommand{\ce}[1]{Eq.~(\ref{#1})}
\newcommand{\cf}[1]{{Fig.~\ref{#1}}}

\newcommand{\eqs}[1]{\begin{equation} \begin{split} #1\end{split} \end{equation} }

\title{Di-$J/\psi$ production at the Tevatron and the LHC}

\ShortTitle{Di-$J/\psi$ production at the Tevatron and the LHC}

\author{\speaker{Jean-Philippe Lansberg}\\
IPNO, Universit\'e Paris-Saclay, Univ. Paris-Sud, CNRS/IN2P3, F-91406, Orsay, France\\
        E-mail: \email{Jean-Philippe.Lansberg@in2p3.fr}}

\author{Hua-Sheng Shao\\
        Theoretical Physics Department, CERN, CH-1211 Geneva 23, Switzerland\\
        E-mail: \email{huasheng.shao@cern.ch}}

\abstract{We briefly review recent results which we have obtained in the study
of $J/\psi$-pair production at the Tevatron and the LHC. 
We claim that the existing data set from CMS and D0 point at a significant
double-parton-scattering contribution with an effective 
cross section smaller than that for jet-related observables.
We have also derived simple relations involving feed-down fractions from excited
states which can help in disentangling the single from the double scatterings.
}

\FullConference{XXIV International Workshop on Deep-Inelastic Scattering and Related Subjects\\
		11-15 April, 2016, 
		DESY Hamburg, Germany}

\begin{document}

\section{Introduction}\vspace*{-0.3cm}

The observation of the associated production of a quarkonium 
with a vector boson or a heavy quark  as well as of a pair of quarkonia is now quasi
the bread and butter of quarkonium physics at the LHC and the Tevatron
with nearly a dozen of experimental analyses~\cite{Aaij:2011yc,Aaij:2012dz,Aad:2014kba,Khachatryan:2014iia,Abazov:2014qba,Aad:2014rua,Aad:2015sda,Aaij:2015wpa,Abazov:2015fbl,ATLAS-CONF-2016-047} accompanied by many relevant theoretical works\footnote{Let us stress here that a number of these theoretical works benefited
from automated tools tailored for quarkonium production. Let us cite {\sc Madonia}~\cite{Artoisenet:2007qm}, {\sc Helac-Onia}~\cite{Shao:2012iz,Shao:2015vga} and FDC~\cite{Wang:2004du}.}  providing predictions before these analysis~\cite{Artoisenet:2007xi,Li:2008ym,Lansberg:2009db,Mao:2011kf,Kom:2011bd,Gang:2012ww,Gang:2012js,Gong:2012ah,Lansberg:2013qka,Lansberg:2013wva,Dunnen:2014eta,Lansberg:2015lva,Likhoded:2016zmk} or interpretations of these results~\cite{Sun:2014gca,Lansberg:2014swa,He:2015qya,Baranov:2015cle,Borschensky:2016nkv,Shao:2016wor,Lansberg:2016rcx}. 
We focus here on $J/\psi$-pair production at the LHC and the Tevatron.

\vspace*{-0.3cm}
\section{$J/\psi$-pair production and the "CMS puzzle"}\vspace*{-0.3cm}

As a matter of fact, $J/\psi$-pair hadroproduction is not a new subject of investigations.
30 years ago,  NA3~\cite{Badier:1982ae,Badier:1985ri} analysed it at the CERN-SPS.
At the LHC, it has been measured by LHCb~\cite{Aaij:2011yc} with an admittedly
small data sample but which covers low $P_T$ and more recently
by the CMS~\cite{Khachatryan:2014iia} and ATLAS~\cite{ATLAS-CONF-2016-047} collaborations 
with a $P_T$ cut of 4 to 8 GeV depending of the rapidity. 
At the Tevatron, the D0 collaboration~\cite{Abazov:2014qba} analysed it
with a $P_T$ cut of 3 GeV.

\begin{figure}[htb!]
\begin{center}\vspace*{-.5cm}
\subfloat{\includegraphics[width=0.42\columnwidth,draft=false]{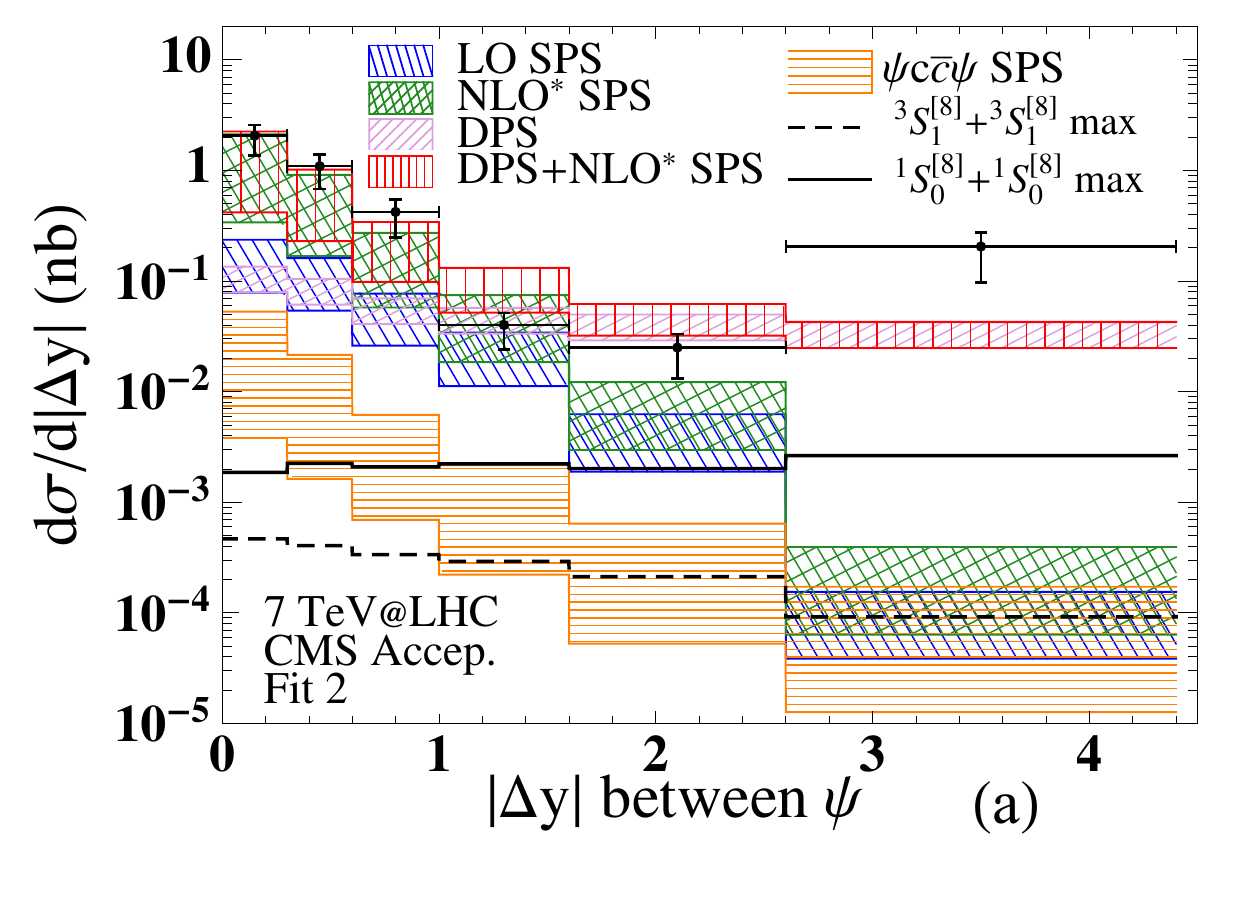}\label{fig:dsigCMSb}}
\subfloat{\includegraphics[width=0.42\columnwidth,draft=false]{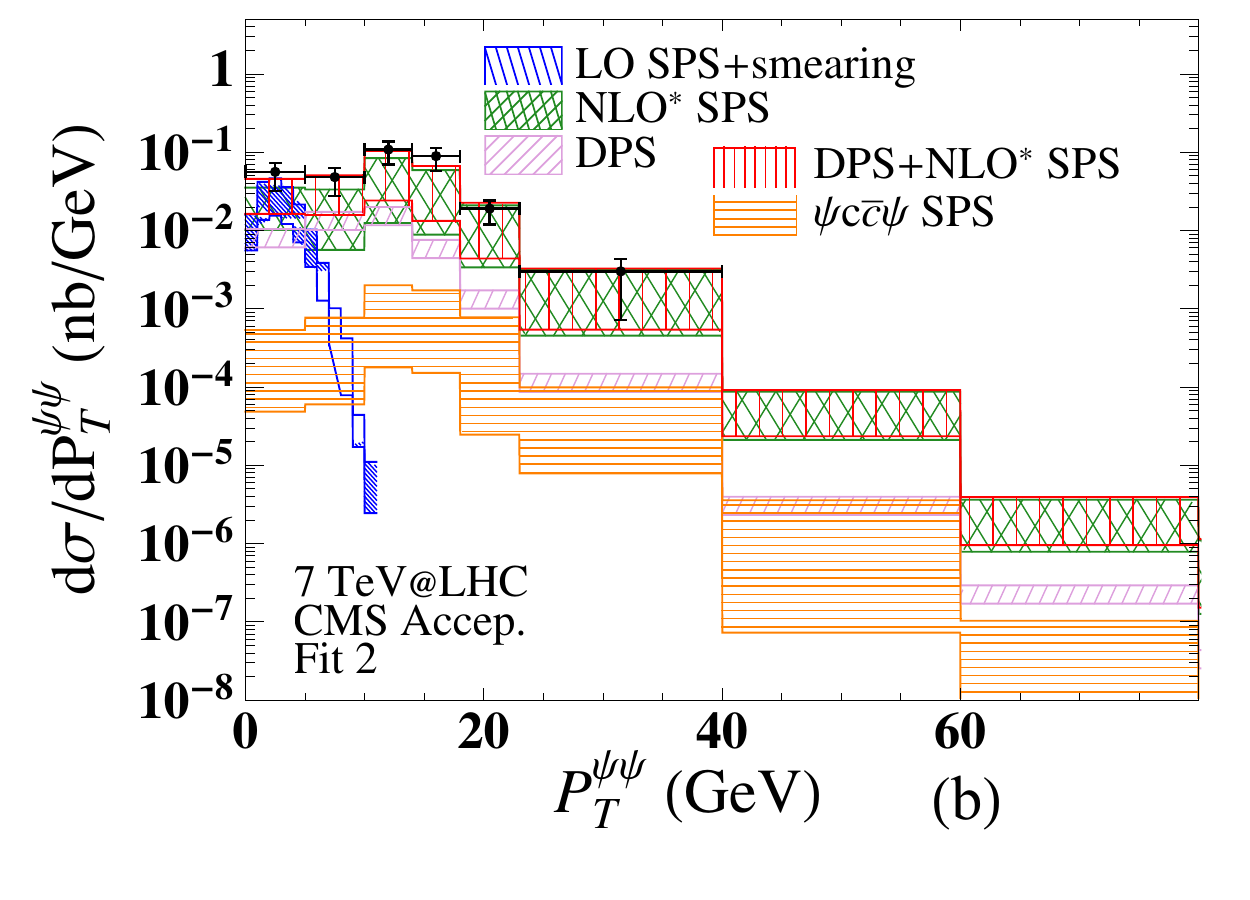}\label{fig:dsigCMSa}}\\
\subfloat{\includegraphics[width=0.42\columnwidth,draft=false]{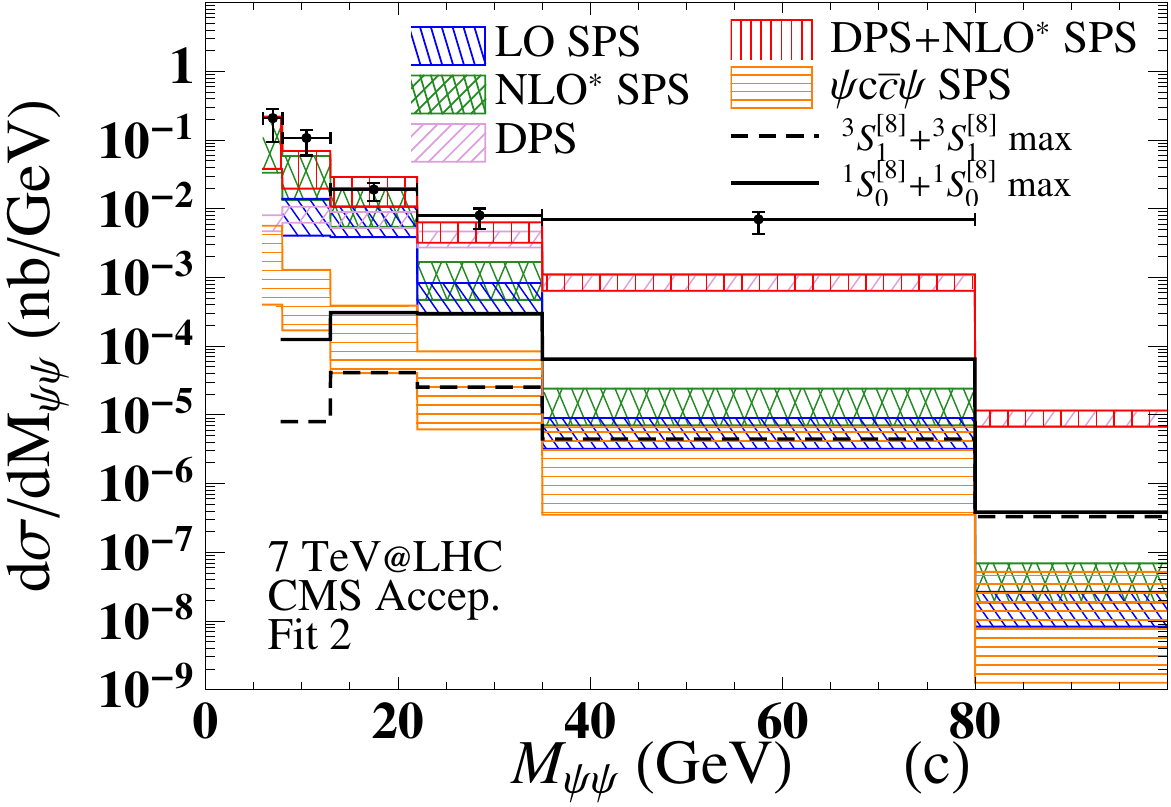}\label{fig:dsigCMSc}}
\subfloat{\includegraphics[width=0.35\columnwidth,draft=false]{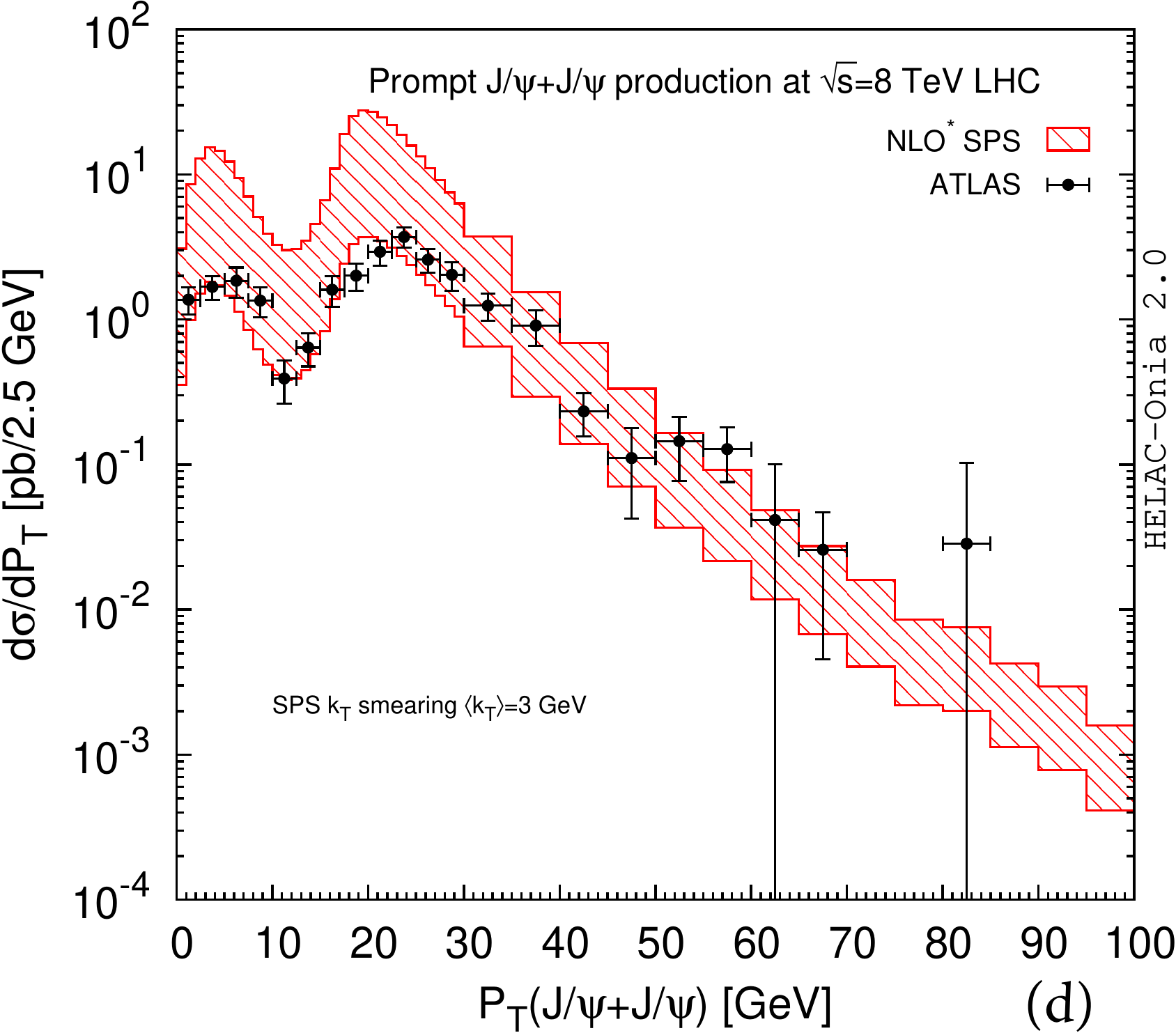}\label{fig:dsigATLAS}}

\caption{Comparisons of different theoretical contributions with the CMS measurement: (a) absolute-rapidity difference ;  (b) pair transverse momentum; (c) pair invariant mass. (d) Idem with the ATLAS data: pair transverse momentum.}
\label{fig:CompareCMS-ATLAS}
\end{center}\vspace*{-.5cm}
\end{figure}

Our claim is that all these data samples are compatible with Colour Singlet (CS) contributions only 
(known up to Next-to-Leading-Order (NLO) accuracy~\cite{Lansberg:2013qka,Sun:2014gca,Likhoded:2016zmk})
at small rapidity separations, $\Delta y$, whereas they point at a significant 
Double Parton Scattering (DPS) contributions for increasing $\Delta y$, compatible with a $\sigma_{\rm eff}$ 
below 10 mb. We guide the reader to \cite{Lansberg:2014swa} for a detailed 
discussion of these different results and to~\cite{He:2015qya,Baranov:2015cle} 
for recent LO NRQCD studies. We find it worth recalling that the D0 and 
ATLAS $J/\psi$-pair analyses~\cite{Abazov:2014qba,ATLAS-CONF-2016-047} are the 
only ones among those of  quarkonium associated production 
(including with a heavy quark or a vector boson) where the DPS and Single Parton Scattering (SPS) contributions 
were separated based on kinematical variables\footnote{We wish to stress that, following
the widespread practice, all the DPS contributions will be understood under
the fully factorised "pocket formula" approach whereby $\sigma^{\rm DPS}_{AB} \propto \sigma_A \sigma _B$. 
This amounts to consider that the parton scattering producing the particle $A$ is completely
uncorrelated with that producing the particle $B$.}.

\cf{fig:CompareCMS-ATLAS} summarises well the situation for data with $P_T$ cuts:
\begin{itemize}\setlength{\itemsep}{-0.3em}
\item the rapidity separation, $\Delta y$, dependence (\cf{fig:dsigCMSb}), agrees very well with
the NLO CS contributions (green band) -- contrary to the LO CS ones (blue band)--
 but for the two last points for $\Delta y \geq 2$. This has been
referred to as the {\it CMS puzzle} and was first discussed in~\cite{Sun:2014gca}.
We claimed in~\cite{Lansberg:2014swa} that the puzzle is very naturally solved
by the inclusion of DPS contributions in an amount compatible with the 
D0 extraction~\cite{Abazov:2014qba} (purple band). Higher QCD corrections at Next-to-Next-to-Leading Order 
(NNLO) are not expected to be significant
(the orange band shows one of the dominant ones). Contrary to the claim made in~\cite{He:2015qya}, 
we do not find the CO contributions relevant here (black lines) unless unphysical\footnote{Not only would they
violate the velocity-scaling rules of NRQCD, they would generate single-$J/\psi$ 
cross sections one or two orders of magnitude larger than all the existing data.}
LDME values are used. If we fit them to this distribution, we obtain 
$\langle {\cal O}^{J/\psi}(^3S_1^{[8]})\rangle = 0.42 \pm 0.12 \hbox{ GeV}^3$ \&
$\langle {\cal O}^{J/\psi}(^1S_0^{[8]})\rangle = 0.91 \pm 0.22  \hbox{ GeV}^3$. Only unexpectedly
large QCD corrections could make these values significantly smaller as to become realistic. 
\item the pair $P_T^{\psi\psi}$ dependence (\cf{fig:dsigCMSa}) is very well accounted by 
the NLO CS contributions.  As expected, the LO contribution cannot account for it
since large-$P_T^{\psi\psi}$ configurations arise from a hard parton with a significant $P_T$ recoiling on
the $J/\psi$ pair\footnote{The impact of the QCD corrections to the $P_T$ spectrum of
quarkonium-pair production was first discussed in~\cite{Lansberg:2013qka}.}. 
A typical $k_T$ smearing of a couple of GeV from initial-state radiations
is not enough to account for the entire spectrum. We note that the CS contributions (as always without
any tuned/fit parameters) agree with the data up to the last bin where $P_T^{\psi\psi}$ is as large as 30 GeV.
The DPS contributions  (purple band) are a bit softer
and are only relevant at low $P_T^{\psi\psi}$. The CO contributions are not shown since they are simply
negligible whatever  the LDME set used is.

\item the invariant-mass dependence (\cf{fig:dsigCMSc}) essentially displays the same
information as the $\Delta y$ dependence. It is normal since, for the majority of 
the events, $M_{\psi \psi}\simeq 2 m_T^{\psi}\cosh{\frac{\Delta y}{2}}$. The inclusion
of the DPS contribution removes the gap with the data in the 4th bin and reduces it
for the last bin which probably contains the exact same events as in the
two last bins of the $\Delta y$ distribution. The maximum allowed CO contributions
are still too low to matter. 

\item \cf{fig:dsigATLAS} shows the corresponding plot of~\cf{fig:dsigCMSa} but for the
ATLAS acceptance and with their preliminary data. The predicted NLO CS contribution (still no tuning) is in good agreement 
with the experimental points.

\end{itemize}

\section{Di-$J/\psi$ production involving feed-down from $\chi_c$ and 
$\psi'$}\vspace*{-0.3cm}

Under the simplistic, yet widely used, approach of the DPS  mechanism
using the so-called pocket formula, one can derive general formulae relating the 
feed-down fractions of the DPS yields for di-$J/\psi$ production to those 
for single-$J/\psi$ production. These are useful 
for two reasons. First, one can employ them  to evaluate the feed-down size and
 thereby improving theoretical predictions. Second, one can also use them
to test a possible hypothesis of DPS-dominance, if hinted at by some typical
kinematical distributions, by measuring the cross section for pair productions 
involving the excited states.

To derive them, one first define specific feed-down fractions for
di-$J/\psi$ inspired from those for single $J/\psi$. These are
$F^{\rm direct}_\psi$, $F^{\chi_c}_\psi$ and $F^{\psi'}_\psi$,  
respectively  for direct production, for production from $\chi_c$ decay or from $\psi'$ decay. 
For $J/\psi+J/\psi$, there are more possibilities. Yet, since it is experimentally challenging
to measure (and subtract) the $\chi_c+\chi_c$ or even $\chi_c+\psi'$ yields, we 
restrict our definition to $F^{\chi_c}_{\psi\psi}$ (resp. $F^{\psi'}_{\psi\psi}$)
as the  $J/\psi+J/\psi$-event fraction from the feed-down of {\it at least} a $\chi_c$ (or resp.  a $\psi'$) decay. 
To phrase it differently, $F^{\chi_c}_{\psi\psi}$ is the fraction of events which include 
a prompt $J/\psi$ (\ie\ direct or from 
$\chi_c$ and $\psi'$ feed-down) plus a $J/\psi$ which is identified as from a $\chi_c$.  
Since it is easier to predict and in spite of being probably 
very difficult to measure, we also define $F^{\rm direct}_{\psi\psi}$ as being the pure  direct 
component, excluding all the possible feed-downs.

Starting from the factorised pocket formula, one easily gets (see \cite{Lansberg:2014swa} for details)
\eqs{\label{eq:FD_fraction_DPS}
F^{\chi_c}_{\psi\psi}= F^{\chi_c}_\psi \times \big(F^{\chi_c}_\psi + 2 F^{\rm direct}_\psi + 2 F^{\psi'}_\psi\big),\,
F^{\psi'}_{\psi\psi}= F^{\psi'}_\psi \times \big(F^{\psi'}_\psi + 2 F^{\rm direct}_\psi + 2 F^{\chi_c}_\psi\big),\,
F^{\rm direct}_{\psi\psi}= (F^{\rm direct}_\psi)^2.}
Using the world average values, 
$F^{\rm direct}_\psi$, $F^{\chi_c}_\psi$ and $F^{\psi'}_\psi$ are close to 60\%, 30\% and 10\% we 
have $F^{\chi_c}_{\psi\psi} \simeq 50 \%$, $F^{\psi'}_{\psi\psi} \simeq 20\%$ 
and $F^{\rm direct}_{\psi\psi} \simeq 35 \%$. 

If SPSs dominate, the feed-downs are significantly different; one expects a larger feed-down from $\psi'$ in the CSM.  
$F^{\psi'}_{\psi\psi}/F^{\rm direct}_{\psi\psi}$  is expected
to be as large as 0.85. As what concerns SPS to $\sigma(\chi_c+J/\psi)$, we have checked that it is indeed suppressed although
it can be kinematically enhanced at large $P_T$~\cite{Likhoded:2016zmk} 
(like $\sigma(J/\psi+\eta_c)$~\cite{Lansberg:2013qka}). $F^{\chi_c}_{\psi\psi}$ should thus be small. 
 In turn, we also have  $F^{\psi'}_{\psi\psi}\simeq 0.85/(1+0.85) \simeq 46 \%$
at any order in $\alpha_s$. $F^{\rm direct}_{\psi\psi}$ should also be close to 55 \%.
We stress that the value of $\sigma_{\rm eff}$ does not appear in \ce{eq:FD_fraction_DPS}.

\begin{figure}[hbt!]
\begin{center}
\includegraphics[width=0.59\columnwidth]{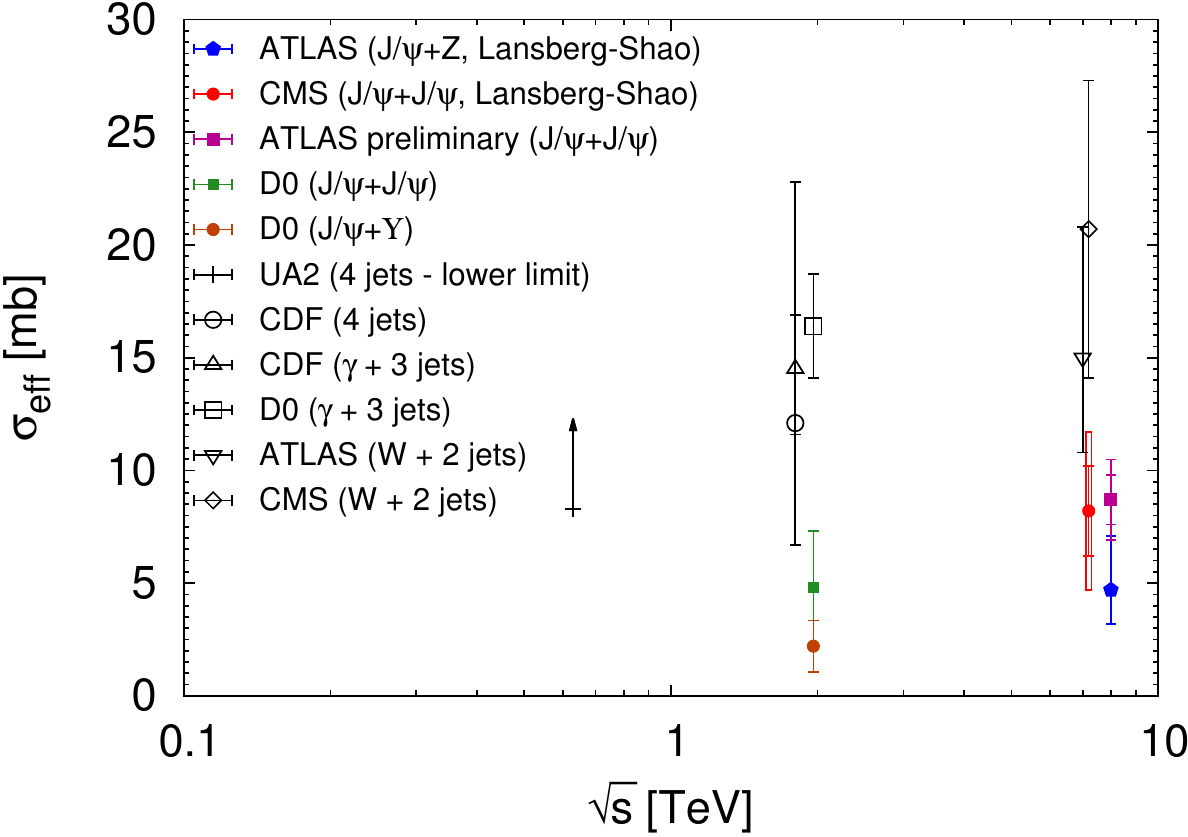}
\caption{
Our ranges for $\sigma_{\rm eff}$  extracted from the $J/\psi+Z$ data 
($4.7^{+2.4}_{-1.5}$~mb)~\cite{Lansberg:2016rcx} 
and from di-$J/\psi$ data~\cite{Lansberg:2014swa} ($8.2\pm 2.0 \pm 2.9$~mb)
compared to other extractions~\protect\cite{Akesson:1986iv,Alitti:1991rd,Abe:1993rv,Abe:1997xk,Abazov:2009gc,Aad:2013bjm,Chatrchyan:2013xxa,Abazov:2014qba,ATLAS-CONF-2016-047}.\vspace*{-1cm}\label{fig:sigma_eff}}
\end{center}
\end{figure}

\section{Conclusion}
\vspace*{-0.5cm}

Many recent experimental studies of associated-production of quarkonia have been lately carried out. 
We have reviewed one of them: $J/\psi$-pair production, for which we have found
that DPS contributions are indispensable with a somewhat small $\sigma_{\rm eff}$ compared
to jet-related observables as illustrated on~\cf{fig:sigma_eff}.
Yet, this value is well within the ballpark of the D0 extraction $J/\psi+\Upsilon$ production and
another we have done from $J/\psi+Z$.

We have derived simple relations for the feed-down fractions from an
excited charmonium state with a $J/\psi$ in the case of the dominance of DPSs, 
which significantly deviate from those for SPSs. Such relations can be used
to disentangle the DPS and SPS regimes.

{\bf Acknowledgements. }
The work of J.P.L. is supported in part by the French CNRS via the LIA FCPPL (Quarkonium4AFTER) and the D\'efi
Inphyniti-Th\'eorie LHC France. H.S.S. is supported by the ERC grant 291377 {\it LHCtheory:
Theoretical predictions and analyses of LHC physics: advancing the precision frontier}.

\bibliographystyle{Science}

\vspace*{-0.3cm}
\bibliography{Onium-DIS2016-021116}

\end{document}